\RequirePackage{fix-cm}
\documentclass[smallextended]{article}       
\usepackage{graphicx}
\usepackage{hyperref}
\usepackage{subfig}
\usepackage{graphicx}
\usepackage{amsmath}
\usepackage{amsfonts}
\usepackage{multirow}
\usepackage{authblk}
\usepackage{booktabs}

\usepackage{tikz}
\usetikzlibrary{arrows,shapes}
\usetikzlibrary{shapes,decorations,arrows,calc,arrows.meta,fit,positioning}
\tikzset{
	-Latex,auto,node distance =1 cm and 1 cm,semithick,
	intervention/.style ={rectangle, draw=black},
	random/.style ={circle, draw=black},
	parameter/.style ={diamond, draw=black},
}

\begin{document}

\title{Inference of large scale relational state processes}

\author[1]{R{\=u}ta Juozaitien{\. e}}
\author[2]{Ernst C. Wit}

\affil[1]{Faculty of Informatics, Vytautas Magnus university, Kaunas, Lithuania
	            \texttt{ruta.juozaitiene@vdu.lt}}
\affil[2]{Università della Svizzera italiana,
	         Lugano, Switzerland \texttt{wite@usi.ch}}

\providecommand{\keywords}[1]
{
	\small	
	\textbf{\textit{Keywords---}} #1
}

\maketitle

\begin{abstract}
Relational states refer to concepts such as friendship or collaboration, in which a relationship persists over a certain amount of time. Study of relational states often involves figuring out what factors contribute to the creation or dissolution of these relationships. However, most methods available now restrict their attention to binary states, i.e., ties that are either present or absent, even though many real-world systems evolve through multiple relational states (e.g., acquaintance, friendship, close friendship). We propose a continuous-time framework for modelling and inferring relational state networks in which each edge evolves by transitioning between two or more states. 

In our model, transition intensities are driven by state-dependent covariates that might be decomposed into anchoring (current-state) and pulling (target-state) mechanisms, with both linear and smooth non-linear effects. We address two common sampling regimes. With full event histories, a Cox-type partial likelihood with nested case–control sampling enables efficient estimation of both parametric and smooth effects. Instead, for panel data we derive a general ODE formulation for the likelihood, which leads to a particularly efficient inference procedure for binary state model. Simulation studies confirm accurate recovery of model parameters, and an empirical application to adolescent friendship data reproduces the substantive conclusions of established modelling techniques while offering substantial computational gains. The framework preserves the interpretability of classical network effects, generalizes them to multi-state ties, and scales to larger, more complex designs under both full-history and panel sampling designs.


\end{abstract}

\section{Introduction}
\label{intro}

Researchers across diverse domains increasingly adopt a dynamic network perspective to study complex phenomena. From patient transfers in healthcare systems \cite{vu2017relational} to the spread of invasive species \cite{boschi2025mixed} or anomalous transactions in financial markets \cite{peng2018anomalous}, such processes can be viewed as time-evolving interactions between nodes — hospitals, regions, or accounts, respectively. Traditional modelling approaches typically rely on binary networks, where nodes represent entities in the system and edges indicate whether a relationship exists between them. Two prominent frameworks in this setting are the stochastic actor-oriented model (SAOM) \cite{snijders2001statistical,snijders2017stochastic,snijders2010introduction} and the temporal exponential random graph model (TERGM) \cite{hanneke2010discrete,leifeld2018temporal}. Both have been widely applied in empirical studies, but they differ in how network evolution is represented. SAOM treats network change as a continuous-time Markov process in which actors are assumed to have full knowledge of the network and make sequential decisions to change their outgoing ties in order to maximize an objective function. This objective function is expressed as a linear combination of network statistics (e.g., reciprocity, transitivity, or degree-related terms), thereby capturing endogenous structural tendencies. TERGM, in contrast, models the probability of observing a network at a discrete time point as a function of network statistics that summarize its current and past configurations. Despite their flexibility, both approaches share important limitations: they are typically restricted to binary ties, meaning that relationships are either present or absent, and thus overlook the heterogeneity and richness of interactions found in real-world systems. Many complex systems involve multiple interaction types or varying relationship strengths, which cannot be adequately captured by a binary framework.

In this paper, we address this limitation by generalizing the binary view to networks in which ties may exist in multiple states. These states represent varying levels or types of relationships (such as acquaintanceship, friendship, or collaboration) and are conceptualized as relatively stable connections, rather than brief, isolated events. This perspective is particularly relevant in the context of multiplex (or multilayer) networks, where multiple layers represent different modes of interaction, social circles, or temporal instances. Multiplex networks preserve the structural and functional distinctions between relationship types, offering a richer and more realistic representation of complex systems. Examples of these systems are prevalent across various domains \cite{su2022evolution,wang2015evolutionary,su2022evolution2}. In social contexts, the same set of individuals may interact across multiple relational layers: personal, professional, and social — both offline and through online platforms (e.g., Facebook, LinkedIn). Conventional approaches often collapse these diverse interactions into a single aggregated tie-type, thereby oversimplifying the complexity of real-world relational structures. For example, information dissemination typically occurs through online networks, whereas disease transmission is constrained to physical contacts. Capturing multiple states of relationships is therefore essential for understanding the true dynamics of complex systems. 

In multi-state networks, ties can take multiple qualitatively distinct forms, such as acquaintanceship, friendship, close friendship, or professional collaboration. These richer state spaces allow ties to evolve not only by forming or breaking, but also by changing in type or intensity. For instance, an individual may transition from being an acquaintance to a friend, or from a friend to a close friend, without the tie ever disappearing. This distinction highlights a key limitation of binary models: all transitions collapse into the same event type — formation or dissolution — thereby ignoring the heterogeneity of relationships. In contrast, multi-state networks capture the qualitative evolution of ties, distinguishing whether a relationship is strengthening, weakening, or shifting across layers of interaction. Such representations are especially important in domains where the nature of the tie, rather than just its presence, drives the dynamics: for example, collaboration versus competition in organizational settings \cite{verschoore2020interplay}, or professional versus personal ties in social systems \cite{becker2020multiplex,han2020building}. 

Most existing research on multiplex or multi-state networks has focused on the static setting, where multiple layers represent different types of relationships among the same set of nodes (e.g., personal vs. professional ties) \cite{kivela2014multilayer, battiston2017new}. In this framework, each layer is a single-mode snapshot, and the goal is often to analyse the structure and interdependence across these layers using diagnostics like community detection, layer coupling, or tensor decomposition \cite{kivela2014multilayer}. While this static framework has proven powerful in applications ranging from transportation \cite{chodrow2016demand} to brain connectivity \cite{de2017multilayer}, it inherently lacks a temporal dimension.

This paper addresses these challenges by introducing a dynamical network framework for relational state analysis. We propose to model the multi-state relational state process as a collection of staggered relational \emph{event} processes, describing the instanteneous state changes.  We consider two data sampling scenarios: (1) a complete setting, where all changes in relational states are recorded, (2) a panel design where a set of individuals is observed at multiple discrete time points, producing cross-sectional network snapshots. In the complete sampling setting, the network dynamics can be inferred using a relational event model. In contrast, the more realistic panel design offers lower temporal resolution, but repeated observations still provide valuable information about tie formation rates and persistence. Analysing such data requires novel inference strategies. To this end, we develop an approach that uses queuing theory to model waiting times between relational state changes, and we embed this formulation within a generalized additive model (GAM) framework. This combination enables flexible incorporation of covariates, thereby linking network dynamics to explanatory mechanisms in a statistically tractable manner. In doing so, our work bridges the gap between transient relational events and persistent relational states, offering a principled, generalizable framework for modelling the evolution of dynamic state networks across diverse domains.

\section{Relational state model}
\label{sec:model}

In this section, we define a relational multi-state process as dynamic graph $G(t)=(V(t),E(t))$ on an temporal interval $t\in [0,T]$. The set of vertices, e.g., individuals, is described as a potentially varying subset of indicators, $V(t)\subset \mathbb{N}$. To describe the multi-state nature of the relationships between the vertices, we define an individual edge \[ E_{sr}(t) = \mbox{state of the tie $(s,r)$ at time $t$},\] with $s,r\in V(t)$ and $E_{sr}(t)\in \mathcal{I}$, the set of possible relationships between $s$ and $r$, i.e., the set $\mathcal{I}$ denotes the finite set of possible tie states. In the simplest case, $\mathcal{I}$ is binary: for example, in a social network where ties represent friendship, a relation between two individuals $s$ and $r$ is either present $E_{sr}=1$ (friendship) or absent $E_{sr}=0$ (no friendship). However, many more complex relationships can be encoded in this way, such as {\bf multi-state relationships}, i.e., $|\mathcal{I}|>2$, or {\bf multiplex relationships}, e.g., $\mathcal{I} = \mathcal{I}_1\times \mathcal{I}_2$.

The stochastic framework to describe the relational state process proposed in this paper builds on the relational event model \cite{bianchi2024relational}, a probabilistic approach describing temporal interactions between a set of nodes $V(t)$. In our setting, a discrete event $e$ at time $t$ corresponds to a change in the state of the relationship between a sender $s \in V(t)$ and receiver $r \in V(t)$. Specifically, the relational transition of the edge $(s,r)$ from state $i$ to state $j$ at time $t$ is represented as a 3-tuple,
\begin{equation}
	\label{eq:multistate}
e = \left((s, r), (i,j), t\right).
\end{equation}
For example, this could refer to the fact that the relationship between persons $s$ and $r$ transitioned at time $t$ from being acquaintances to becoming friends. In this setting, an event corresponds to the formation or dissolution of a tie, and  network dynamics reduce to tracking the appearance and disappearance of edges. We do not explicitly model the appearances or disappearances of the vertices in the vertex set. This is assumed part of the information given in the underlying filtration $\mathcal{F}$. 

We propose a modeling the transitions of the muti-state relational state process via a multivariate counting process $N=\{N_{s,r}^{(i,j)}\}$, whose components describe
\[ N_{sr}^{(i,j)}(t) = \mbox{$\#$ transitions of edge $(s,r)$ from states $i$ to $j$ until time $t$.} \]
We assume that event times are distinct, so no two edges change state simultaneously. In addition, each counting process $N_{sr}^{(i,j)}$ is assumed to have finite expectation, satisfy the Markov property, and have no events at time 0. Under these conditions, the process $N_{sr}^{(i,j)}$ is an adapted, non-negative, and non-decreasing process with finite expectation, and, therefore, a submartingale. The Doob–Meyer decomposition then guarantees that the counting process can be expressed as the sum of a predictable component $\Lambda_{sr}^{(i,j)}(t)$, capturing the systematic dynamics, and a martingale component $M_{sr}^{(i,j)}(t)$ representing random fluctuations,
\begin{equation*}
	N_{sr}^{(i,j)}(t) = M_{sr}^{(i,j)}(t) + \Lambda_{sr}^{(i,j)}(t).
\end{equation*}
If the distribution of waiting times between two transitions is absolutely continuous, the predictable component admits the integral representation,	
\begin{equation*}
	\Lambda_{sr}^{(i,j)}(t) = \int_0^t \lambda_{sr}^{(i,j)}(\tau)~\mathrm{d}\tau,
\end{equation*}
where $\lambda_{sr}^{(i,j)}(t)$ is the intensity function, describing the instantaneous rate at which the relation between $s$ and $r$ transitions from state $i$ to state $j$ at time $t$. In this paper, we propose to model the intensity via an exponential hazard \cite{cox1972regression},
\begin{equation}
	\label{eq:hazard}
	\lambda_{sr}^{(i,j)}(t) = Y_{sr}^{(i,j)}(t) \lambda_0^{(i,j)}(t) \exp\left[ f^{(i,j)}(x_{sr}^{\,{(i,j)}}(t)) \right], i \neq j,
\end{equation}
where $Y_{sr}^{(i,j)}(t)$ is a binary risk indicator equal to 1 if edge $(s,r)$ is at risk of transitioning rom $i$ to $j$ at time $t$, and 0 otherwise. For example, if $(s,r)$ is simply not in state $i$ at time $t$, then it is not at risk of transitioning from $i$. $\lambda_{0}^{(i,j)}(t)$ is the baseline hazard function for the transition $i \rightarrow j$. The vector $x_{sr}^{\,{(i,j)}}(t)$ is a collection of endogenous and exogenous, possibly time-varying, covariates, whose transition-specific effects are described by the function $f^{(i,j)}$. One special case involves a linear specification $f^{(i,j)}(x_{sr}^{\,{(i,j)}}(t)) = \beta^{\,{(i,j)}^\top} x_{sr}^{\,{(i,j)}}(t)$, where $\beta^{\,{(i,j)}}$ are the corresponding effect sizes.

\subsection{State-dependent network covariates}

The specification of model covariates is one of the central aspects of defining the intensity function. Exogenous attributes, such as node-level covariates (e.g., age, gender) or dyadic covariates (e.g., geographic distance), can usually be incorporated directly. However, greater care is required in defining endogenous network effects, which capture internal dependencies in the evolving network. In the traditional binary case, such effects are typically straightforward: for instances, reciprocity simply indicates whether the reverse edge $(r,s)$ is present, while transitivity counts the number of two-paths between $s$ and $r$. In the multi-state setting, ties can vary in strength or type, and these distinctions may influence the dynamics of transitions. For example, a reciprocated strong tie may shape future interactions differently than a reciprocated weak tie, and a triad closed by weak links may not carry the same structural significance as one closed by strong links. To account for these nuances,  endogenous covariates must be generalized to incorporate tie-state information, thereby giving rise to state-dependent network covariates. This generalization ensures that structural mechanisms remain meaningful in settings where ties evolve beyond the binary present/absent framework.
	 
\paragraph{Multi-state reciprocity.} To illustrate the idea of state-dependent network effects, we begin with reciprocity, one of the most fundamental mechanisms shaping relational dynamics. It reflects the idea that individuals tend to respond to an action with an equivalent action: they invest in social bonds with the expectation of receiving benefits in return. When such expectations are met, reciprocal ties reinforce mutual trust and strengthen the relationship \cite{friedkin2004social}. In the binary case, reciprocity is typically defined as an indicator of whether the reverse tie is present. In a multi-state setting, however, reciprocity does not simply reflect whether a tie is returned, but also \textit{how} it is returned. When ties can differ in form or strength, the meaning of reciprocity depends on the state of the reverse tie. For example, if individual $s$ considers individual $r$ as a close friend, while $r$ regards  $s$ only as an acquaintance, the relationship is technically reciprocated but with unequal intensity. To capture this nuance, we define the state-dependent reciprocity covariate as the indictor function
\begin{equation*}
	\text{Rec}^{(j)}_{sr}(t)  = \mathbf{1}_{\{ E_{rs}(t) = j\}}, \quad j \in \mathcal{I}.
\end{equation*}
This covariate can capture the tendency for the tie $(s,r)$ to transition from its current state of the reverse tie $(r,s)$. In other words, reciprocity reflects the tendency of a tie to align with its reverse tie by moving into the same state.

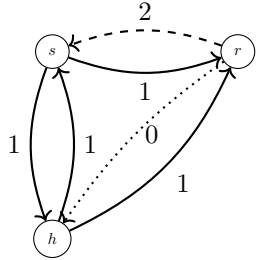
\begin{figure}[ht]
	\centering
\begin{tikzpicture}[baseline=(current bounding box.north)]
	\node[random, scale=0.6] (n1) {~$s$~~};
	\node[random, scale=0.6] (n2) [right=2cm of n1] {~$r$~~};
	\node[random, scale=0.6] (n3) [below=2cm of n1, align=center] {~$h$~};
	
	\path[draw,thick,->]
	(n1) edge [bend right=20] node[left] {1} (n3)
	(n1) edge [bend right=20] node[below] {1} (n2)
	(n3) edge [bend right=20] node[below right] {1} (n2)
	(n2) edge [bend right=10, dotted] node[right] {0} (n3)
	(n2) edge [bend right=20, dashed] node[above] {2} (n1)
	(n3) edge [bend right=20] node[right] {1} (n1);
 \end{tikzpicture}
\caption{Illustration of state-dependent reciprocity in a multi-state network. Solid, dashed, and dotted arrows represent ties in different states.}
\label{fig:reciprocity_example}
\end{figure}

To illustrate state-dependent positive reciprocity effect $\beta_{\mbox{\scriptsize rec}}^{(i,j)}>0$, consider the example in Figure ~\ref{fig:reciprocity_example}, where nodes $s$, $r$ and $k$ are connected by ties in different states. Suppose we are interested in the transition intensity of the tie $(s,r)$. Since the reverse edge $(r,s)$ is in state 2 (dashed arrow), it increases the rate that  $(s,r)$ transitions to a higher state. At the same time, the tie $(r,s)$ is itself influenced by the fact that $(s,r)$ is currently in state 1 (solid arrow), which promotes a transition of $(r,s)$ toward state 1. In a binary framework, both ties would simply be treated as reciprocated, whereas the state-dependent formulation distinguishes between the type of reciprocation involved.

\paragraph{Multi-state triadic closure.} Another commonly observed feature of social interactions is the tendency to create closed relational structures. Such patterns often emerge through path-shortening, where new ties are more likely to form between actors who already share mutual contacts. It is often assumed that individuals experience normative or psychological pressure from their direct contacts, encouraging them to establish connections that preserve balance and coherence within their social environment \cite{giuliani2008industrial}. Triadic closure may arise through several mechanisms, including transitivity, sender or receiver balance, and cyclic closure \cite{robins2009closure}. The basic idea is that the probability of a tie between two nodes increases with the number of two-paths connecting them. 

For example, a state-dependent transitive closure covariate can be defined as
\begin{equation*}
	\text{Tr}^{(i)}_{sr}(t)  = \sum_{h \in V(t)\backslash \{s,r\}} \mathbf{1}_{\{ E_{sh}(t) =i= E_{hr}(t) \}}, \quad i \in \mathcal{I},
\end{equation*}
which counts the number of two-paths from $s$ to $r$ where the intermediate edges are both in state $i$. This state-dependent formulation can be directly extended to other types of triadic closure, allowing their influence on tie dynamics to depend on the qualitative form of the relationships involved.

\begin{figure}[ht]
	\centering
	\begin{tikzpicture}[baseline=(current bounding box.north)]
	\node[random, scale=0.6] (n1) {~$s$~};
	\node[random, scale=0.6] (n2) [right=2cm of n1] {~$r$~};
	\path (n1) -- (n2) coordinate[midway] (mid);
	\node[random, scale=0.6] (n3) [above=2cm of mid, align=center] {~$h$~};
	
	\path[draw,thick,->]
	(n1) edge [bend right=20] node[left] {1} (n3)
	(n1) edge [bend right=20, dotted] node[below] {0} (n2)
	(n3) edge [bend right=20] node[right] {1} (n2)
	(n2) edge [bend right=10, dotted] node[right] {0} (n3)
	(n2) edge [bend right=20, dashed] node[above] {2} (n1)
	(n3) edge [bend right=20] node[left] {1} (n1);
\end{tikzpicture}
\caption{Illustration of state-dependent transitivity in a multi-state network.}
\label{fig:transitivity_example}
\end{figure}
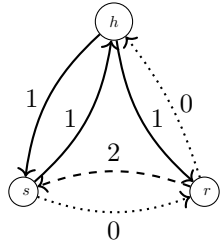

As an illustration of positive state-dependent transitivity effect $\beta_{\mbox{\scriptsize tr}}^{(i,j)}>0$, consider Figure~\ref{fig:transitivity_example}, where nodes $s$, $r$ and $h$ form a potential triad. The ties $(s,h)$ and $(h,r)$ are both in state 1 (solid arrows), creating a two-path from $s$ to $r$. Consequently, the edge $(s,r)$ is attracted to state 1 by transitivity. At the same time, the edge $(s,h)$ is subject to a transitivity effect of type 0, since there exists a two-path $s \rightarrow r \rightarrow h$ in state 0 (dotted arrows). Not all types of transitivity might be equally influential: two-paths composed of strong ties (e.g., state 2) may provide a more compelling basis for closure than those composed of weaker ties (e.g., state 0). This example demonstrates how state-dependent transitivity allows to distinguish between closures formed through ties in different states. 

\paragraph{Other state-dependent endogenous covariates.} Beyond reciprocity and triadicity, a wide range of network covariates have been considered in relational event and stochastic actor-oriented models, including degree-related statistics (e.g., in-degree and out-degree), dyadic-level statistics (e.g., turn-taking, turn-continuing) and path-based statistics (e.g., $l$-path connectivity). Most of these can be adapted to the multi-state framework. For example, degree-related measures can be generalized by counting the number of incoming or outgoing ties that are currently in a given state. The out-degree of node $s$ in state $i$ measures how many outgoing ties from $s$ currently occupy that state:
\begin{equation*}
	\text{OutDeg}^{(i)}_{s}(t) = 
	\sum_{r \in V(t) \setminus \{s\}} \mathbf{1}_{\{\, E_{sr}(t) = i \,\}},
	\quad i \in \mathcal{I}.
\end{equation*}
Analogously, the in-degree of node $r$ in state $i$ counts the number of incoming ties that point to $r$ and are presently in state $i$:
\begin{equation*}
	\text{InDeg}^{(i)}_{r}(t) = 
	\sum_{s \in V \setminus \{r\}} \mathbf{1}_{\{\, E_{sr}(t) = i \,\}},
	\quad i \in \mathcal{I}.
\end{equation*}
These measures capture heterogeneity in activity and popularity across nodes, extending the classical degree concepts to networks where ties can assume multiple states.

Although we introduced endogenous network effects through indicator-based definitions, they can also be expressed in alternative ways, ranging from simple binary measures to more elaborate temporal functions that account for the timing between state changes \cite{juozaitiene2024s}. However, not all of these alternatives remain meaningful in the multi-state setting. For example, some formulations of reciprocity define it as a cumulative quantity, representing the total volume of past relational events from $r$ to $s$ up to time $t$. Within a state-based framework, such a definition is difficult to interpret: if a reverse edge repeatedly alternates between states, the cumulative count of changes offers little insight into the likelihood that the tie of interest will align with it. What matters instead is the current state of the reverse tie. It is more natural to assume that the probability of a transition depends on whether the reverse tie presently occupies a given state, rather than on the number of times it has shifted in the past. Nevertheless, temporal variations of endogenous network effects can still be introduced, but they require careful reformulation. In the case of reciprocity, one could define a covariate measuring the time elapsed since the reverse tie last changed its state. This would capture typical reaction time until alignment with the reverse tie occurs. However, such temporal definitions require the complete state-change history of a network, as they depend on the precise transition times. In panel designs, where only discrete snapshots are observed, this information is not available, making these temporal formulations less practical.

\subsection{Parameter constraints for increased interpretability}
\label{subsec:parameters}

Parameter restrictions play an important role in defining the intensity function, as they help control model complexity and reflect substantive assumptions regarding network dynamics. In the general model specification \eqref{eq:hazard}, every possible transition $i \rightarrow j$ is associated with its own linear effect $\beta^{\,{(i,j)}}$ or non-linear effect $f^{\,{(i,j)}}$. This formulation allows covariates to influence transitions in a state-specific manner. For example, when one individual considers another a friend, reciprocity can substantially increase the likelihood that the second person will reciprocate by deepening the tie from acquaintance to friendship. However, once two individuals are already friends, reciprocity may operate differently. If one person elevates the relationship to the next level, i.e., regarding the other as a close friend, the other may not feel the same obligation to reciprocate with the same intensity as before. In this case, reciprocity exerts a weaker influence on the transition from friendship-to-close-friendship compared to its effect in the acquaintance-to-friend transition. More generally, reciprocity (and other endogenous mechanisms) need not act as uniform drivers of relational change, as their effects may be inherently state-dependent, potentially being stronger in early stages of tie formation and weaker as ties evolve to higher states, or vice versa. 

\paragraph{SAOM-type restrictions.} While this specification provides a rich and flexible representation of tie dynamics, it also leads to a potentially large number of parameters, especially when the state space $\mathcal{I}$ is large. In practice, it is often desirable to impose restrictions that reduce complexity while reflecting plausible modelling assumptions. A well-known example comes from stochastic actor–oriented models (SAOM) \cite{snijders2001statistical}, where tie creation and dissolution are assumed to have symmetric effects. Substantively, this implies that forming a new tie and dissolving an existing one require the same effort, with effects equal in magnitude but opposite in sign, i.e. $\beta^{(0,1)} = -\beta^{(1,0)}$. Analogous constraints can also be imposed in the multi-state setting, for instance by requiring $$\beta^{(i,j)} = -\beta^{(j,i)}.$$
This type of effect-symmetry can be sensible way to cut down the number of parameters by a factor of two. However, there may be circumstances in which this is unlikely. For example, it may be that creating a friendship is easier than dissolving one. In this case, there may be other types of parameter restrictions to manage model complexity while preserving substantive interpretation. 

\paragraph{Anchor-and-pull restrictions.}
We propose a set of restrictions in which transition intensities depend simultaneously on the current state $i$ and the target state $j$. The intuition is that transitions are shaped both by the forces that keep a tie anchored in its current state and by factors that pull it toward the target state. Formally, the linear predictor of the endogenous network effect $x$ in the intensity function can be written as a combination of $x$-covariates associated with the two states:
\begin{equation}
	\label{eq:restrictions}
	 \mbox{LP}_{sr}^{(i,j)}(x)=\beta^{+}_x x_{sr}^{(j)}(t) + \beta^{-}_x x_{sr}^{(i)}(t), \quad i,j \in \mathcal{I},
\end{equation}
where $x_{sr}^{(j)}(t)$ is the $x$-covariate defined from the perspective of the target state $j$, representing mechanisms that encourage the transition, while $ x_{sr}^{(i)}$ is the same covariate defined from the perspective of the current state $i$, representing forces that sustain the tie in its present state $i$. The corresponding parameter vectors $\beta^{+}$ and $\beta^{-}$ capture, respectively, the effect of mechanisms pulling a tie toward the target state and those anchoring it in the current state.

To illustrate, consider a network with three states $\{0,1,2\}$ and a tie $(s,r)$ currently in state 0. If we focus on the rate of transition $0 \rightarrow 1$. Reciprocity may influence this intensity in two distinct ways. If the reverse tie $(r,s)$ is in state 1, reciprocity of type 1 increases the likelihood of the transition, since moving to state 1 would result in a reciprocated relation. Conversely, if the reverse tie is in state 0, reciprocity of type 0 may encourage  $(s,r)$ to remain in its current state rather than transitioning upward. Reciprocity of type 2 plays no role in this transition: even if $(r,s)$ is in state 2, that information s not directly relevant when choosing between staying in state 0 and moving to state 1. The same reasoning extends to triadic effects. For instance, transitivity of type 1 can be seen as a factor pulling the tie toward state 1, by increasing the likelihood of closure through two-paths in state 1. Transitivity of type 0, by contrast, reflects a force that reinforces staying in state 0, while transitivity of type 2 is irrelevant for the $0 \rightarrow 1$ transition.
	
The specification of model covariates can be extended beyond fixed linear effects. In particular, coefficients can be allowed to vary smoothly with time or other continuous variables, thereby capturing non-linear or time-varying influences. Instead of assuming constant effects $\beta^{-}$ and $\beta^{+}$ for anchoring and pulling mechanisms respectively, the linear predictors for the $x$-covariate can be modelled as smooth functions, 
\begin{equation*}
	 \mbox{LP}_{sr}^{(i,j)}(x)=f^{+}_x\big(x_{sr}^{(j)}(t)\big) + f^{-}_x\big(x_{sr}^{(i)}(t)\big),\quad i,j \in \mathcal{I},
\end{equation*}
where $f^{+}_x(\cdot)$ and $f^{-}_x(\cdot)$ are spline-based functions of covariates defined from the perspectives of the target and current states. This formulation enables the intensity function to capture patterns such as non-linear dependencies, duration effects, or gradual temporal changes that cannot be represented by fixed coefficients.

\section{Inference}

For fitting the above relational state networks, we consider two data sampling scenarios. In the first scenario, we consider a complete data setting, where every transition of an edge from one state to another is recorded together with its exact timing. This setting provides the most detailed view of network evolution and allows the process to be represented as a collection of staggered relational event models. In the second setting, we consider a panel design, where a fixed set of nodes is observed at several fixed time points, producing cross-sectional network snapshots. Compared to the complete event design, the panel setting is more common in practice but also more challenging: the exact ordering of transitions between observation points is unobserved, and multiple unrecorded events may occur in between. This loss of temporal resolution necessitates the development of new inference strategies tailored to panel data, ensuring that meaningful estimates of transition dynamics can still be obtained.

\subsection{Complete data setting}

In the complete setting, the full temporal history of network evolution is observed. For every edge, we record the exact moments at which state transitions occur. Formally, the data take the form of the temporal sequence of state-transitions $(e_1, \ldots,e_n)$, where each transition $e_k$ of edge $(s_k,r_k)$ from state $i_k$ to state $j_k$ at time $t_k$ is defined according to \eqref{eq:multistate}. At the initial time $t=0$, the network configuration is fully observed: each dyad $(s,r)$ occupies a known starting state $E_{sr}(0) \in \mathcal{I}$, which may differ across edges. The subsequent network evolution is then captured by an ordered sequence of $n$ transitions 
\begin{equation*}
	\{(s_k, r_k), (i_k,j_k), t_k)~|~ k =1, \ldots, n\}, ~~0 < t_1 < \ldots t_n \leq T, 
\end{equation*} 
where each event specifies the sender, receiver, initial and new state, and the exact transition time.

Following \cite{cox1972regression}, the effect of model covariates on transition intensities (see equation~\eqref{eq:hazard}) can be estimated using a partial likelihood approach. This method conditions on the risk set$\mathcal{R}(t_k)$ at each event time $t_k$, which consists of all dyads-state transitions that were at risk of occurring at that instant. The contribution of event $k$ to the likelihood is the intensity of the observed transition divided by the sum of intensities across the entire risk set, 
	\begin{equation*}
	L^P(f)  =\prod_{k=1}^n \frac{\lambda_{s_kr_k}^{(i_k,j_k)}(t_k)}{ 
		\displaystyle\sum_{\substack{((s,r),(i,j))\in \mathcal{R}(t_k)}} \lambda_{sr}^{(i,j)}(t_k)}  
\end{equation*}
where the risk set $\mathcal{R}(t_k)$ consists of tuples $((s,r),(i,j))$ of ties $(s,r)$ that are at risk of transitioning from state $i$ to $j$ at time $t_k$. 
This formulation generalizes the Cox partial likelihood to the setting of multi-state relational events. 

In many empirical applications, large networks generate vast risk sets, whose size typically grows quadratically with the number of nodes \cite{lerner2020reliability}. Using the partial likelihood under such conditions quickly becomes computationally infeasible. A strategy to mitigate this complexity is nested case-control sampling \cite{borgan1995methods,vu2015relational}, which reduces the size of the risk set by retaining the observed event together with a small random sample of non-transition drawn from the full risk set. 

\paragraph{Linear model with common baseline hazard.} Using a sample of a single random non-transition $(s_k^\ast,r_k^\ast)$ from $i_k^\ast$ to $j_k^\ast$, the resulting partial likelihood using a linear model specification with a common baseline hazard $\lambda_0^{(i,j)}=\lambda_0$ is
\begin{eqnarray*}
	PL_{NCC}(\beta)  &=&\prod_{k=1}^n \frac{\lambda_{s_kr_k}^{(i_k,j_k)}(t_k)}{ \lambda_{s_kr_k}^{(i_k,j_k)}(t_k) + \lambda_{s_k^\ast r_k^\ast}^{(i_k^\ast,j_k^\ast)}(t_k)}\\
	&=&\prod_{k=1}^n \frac{\exp\Big(\beta^\top  \Delta x_k\Big)}{ 1 +\exp\Big(\beta^\top  \Delta x_k\Big)},
\end{eqnarray*}
which simplifies to the likelihood of a degenerate logistic regression, where $\beta=\{ \beta^{(i,j)} \}_{ij}$ collects all state-dependent effect parameters and the vector $\Delta x_k$ is the covariate contrast for event $k$, defined as
\[ \Delta x_k^{(i,j)} =\left\{ 
       \begin{array}{rl} 
       	0 & \mbox{for } i_k \neq i \neq i_k^\ast \mbox{ and } j_k \neq j \neq j_k^\ast\\
       	x^{(i_k,j_k)}_{s_k r_k}(t_k) & \mbox{for } i=i_k \mbox{ and } j=j_k\\ 
       	- x^{(i_k,j_k)}_{s^\ast_k r^\ast_k}(t_k) & \mbox{for } i=i_k^\ast \mbox{ and } j=j_k^\ast\\ 
       	x^{(i_k,j_k)}_{s_k r_k}(t_k)- x^{(i_k,j_k)}_{s^\ast_k r^\ast_k}(t_k) & \mbox{if } i=i_k=i_k^\ast \mbox{ and } j=j_k=j_k^\ast
       	\end{array} \right.\]
Inference of the effects $\beta=\{ \beta^{(i,j)} \}_{ij}$ can be done by stacking these quantities $\Delta x_k$ ($k=1,\ldots,n$) into a design matrix and fitting logistic regression with $n$ successes and without an intercept. 

\paragraph{General additive mixed model.} Inference can easily be extended to non-linear additive mixed models without a common baseline hazards assumption. The state-dependent baseline hazard can be included into the exponent as a non-linear additive function of time, and the additive mixed effects can be fitted via a penalized degenerate logistic regression \cite{wood2017generalized,boschi2026introduction}.

\subsection{Panel data setting}

In many empirical applications, a complete history of state changes is not available. Instead, networks are observed only at discrete time points, producing panel data in the form of cross-sectional snapshots. To adapt the continuous-time framework to such settings, we model transitions between tie states across observation intervals. These transitions can be expressed in terms of the underlying waiting times until a tie changes state, which naturally leads to a formulation based on competing exponential clocks inspired by queuing theory. 

In a continuous-time Markov chain (CTMC), the waiting time in state $i$ is distributed a generalized exponential with its rate equal to the total hazard of leaving state $i$. Under the constant rate assumption, the dynamics for each edge $(s,r)$ with finite state space $\mathcal{I}$ is governed by the transition-rate matrix
\begin{equation*}
Q_{sr}(t) = (q_{ij}(t))_{i,j \in \mathcal{I}},
\end{equation*}
whose off-diagonal elements $q_{ij}(t) = \lambda_{sr}^{(i,j)}(t), i \neq j$ represent the rate at which edge $(s,r)$ moves from state $i$ to $j$ at time $t$. The diagonal elements satisfy
\begin{equation*}
	q_{ii}(t) = -\sum_{j \neq i} \lambda_{sr}^{(i,j)}(t),
\end{equation*}
ensuring that each row of $Q_{sr}(t)$ sums to zero. The exact transition probabilities over an interval of length $\Delta t_k = t_{k+1}-t_k$ are obtained from the matrix exponential,
\begin{equation*}
P_{\Delta t_k} = \exp{\Big(\int_{t_k}^{t_{k+1}} Q_{sr}(t)~dt\Big)}.
\end{equation*}
In general state spaces this expression is difficult to compute, since evaluating matrix exponentials is computationally demanding and grows rapidly in complexity with the number of states. A common simplification is therefore to assume that transition intensities are piecewise constant within each observation interval. Under this assumption, transition probabilities can be approximated in short intervals as
\begin{equation*}
	P\bigg(E_{sr}(t_{k+1}) = j \;\Big|\; E_{sr}(t_k) = i \bigg) \approx \lambda_{sr}^{(i,j)}(t_k) \Delta t_k + o( \Delta t_k), \quad i \neq j,
\end{equation*}
which shows that the probability of moving from $i$ to $j$ is proportional to the corresponding transition intensity. For $i=j$, the probability of remaining in the same state requires aggregating over all possible exits from state $i$:
\begin{equation*}
	P\bigg(E_{sr}(t_{k+1}) = i \;\Big|\; E_{sr}(t_k) = i \bigg) \approx 1- \sum_{j \neq i}\lambda_{sr}^{(i,j)}(t_k) \Delta t_k + o( \Delta t_k).
\end{equation*}
This aggregation of many competing processes complicates interpretation and estimation. Combined with the computational burden of large transition matrices, this makes estimation in multi-state panel settings substantially more challenging. To keep the framework analytically tractable and directly comparable with existing approaches, we therefore restrict attention to the binary case.

\subsubsection{Two-States Scenario}
\label{subsubsec:two_state}

Consider the special case $\mathcal{I} = \left\{0,1\right\}$, where each edge $(s,r)$ is either absent (state 0) or present (state 1). Transitions are governed by two processes: tie formation with rate $\lambda_{sr}^{(0,1)}(t)$ and tie dissolution with rate $\lambda_{sr}^{(1,0)}(t)$. The dynamics can be represented by the transition-rate matrix of the continuous-time Markov chain: 
\begin{equation*}
	Q_{sr}(t) = \begin{pmatrix}
- \lambda_{sr}^{(0,1)}(t) & \lambda_{sr}^{(0,1)}(t)\\
\lambda_{sr}^{(1,0)}(t) & -\lambda_{sr}^{(1,0)}(t).
	\end{pmatrix}
\end{equation*}
Because the system has only two states, the matrix exponential can be computed explicitly, giving probability of transitioning from state $i$ to state $j$ within $\left[t_k,t_{k+1}\right)$ is
\begin{align*}
		P\bigg(E_{sr}(t_{k+1}) = j \;\Big|\; E_{sr}(t_k) = i \bigg) = &\frac{\lambda_{sr}^{(i,j)}(t_k)}{\lambda_{sr}^{(0,1)}(t_k) + \lambda_{sr}^{(1,0)}(t_k)} \times \\
		&\times \bigg(1-\exp{\Big[-(\lambda_{sr}^{(0,1)}(t_k) + \lambda_{sr}^{(1,0)}(t_k))\Delta t_k\Big]} \bigg), 
\end{align*}
for $i \neq j$. Conversely, the probability of remaining in the same state is
\begin{align*}
	P\bigg(E_{sr}(t_{k+1}) = i \;\Big|\; E_{sr}(t_k) = i \bigg) = & 1- \frac{\lambda_{sr}^{(i,j)}(t_k)}{\lambda_{sr}^{(0,1)}(t_k) + \lambda_{sr}^{(1,0)}(t_k)} \times \\
	&\times \bigg(1-\exp{\Big[-(\lambda_{sr}^{(0,1)}(t_k) + \lambda_{sr}^{(1,0)}(t_k))\Delta t_k\Big]} \bigg).
\end{align*}
Multiplying these contributions across all dyads and observation intervals $k= 1, \ldots, n-1$ gives the full likelihood
\begin{align}
L(\boldsymbol{\beta})  = \prod_{k=1}^{n-1} \Bigg[&\prod_{\substack{sr: \\ 0 \to 1}}\frac{\lambda_{sr}^{(0,1)}(t_k)}{\lambda_{sr}^{(0,1)}(t_k)+\lambda_{sr}^{(1,0)}(t_k)}\bigg(1-\exp\Big[-(\lambda_{sr}^{(0,1)}(t_k)+\lambda_{sr}^{(1,0)}(t_k))\Delta t_k\Big]\bigg) \nonumber \\ 
\times &	\prod_{\substack{sr: \\ 0 \to 0}} 1- \frac{\lambda_{sr}^{(0,1)}(t_k)}{\lambda_{sr}^{(0,1)}(t_k)+\lambda_{sr}^{(1,0)}(t_k)}\bigg(1-\exp{\Big[-(\lambda_{sr}^{(0,1)}(t_k)+\lambda_{sr}^{(1,0)}(t_k))\Delta t_k\Big]}\bigg)\nonumber\\
\times&	\prod_{\substack{sr: \\ 1 \to 0}} \frac{\lambda_{sr}^{(1,0)}(t_k)}{\lambda_{sr}^{(0,1)}(t_k)+\lambda_{sr}^{(1,0)}(t_k)}\Big(1-\exp{\bigg[-(\lambda_{sr}^{(0,1)}(t_k)+\lambda_{sr}^{(1,0)}(t_k))\Delta t_k\Big]}\Bigg)  \nonumber\\
\times&	\prod_{\substack{sr: \\ 1 \to 1}} 1- \frac{\lambda_{sr}^{(1,0)}(t_k)}{\lambda_{sr}^{(0,1)}(t_k)+\lambda_{sr}^{(1,0)}(t_k)}\bigg(1-\exp{\Big[-(\lambda_{sr}^{(0,1)}(t_k)+\lambda_{sr}^{(1,0)}(t_k))\Delta t_k\Big]}\bigg) \Bigg], \label{eq:lik}
\end{align}
where each product corresponds to one of the four possible observed transitions 
($0\to 1, 0\to0,  1\to0, 1 \to 1$) for each tie, ensuring that the model accounts for both changes and persistence of ties.
These probabilities can be interpreted more intuitively by splitting them into two parts. First, the ratio of intensities reflects the relative likelihood of tie formation versus dissolution. Because the intensities follow an exponential form (see eq. \eqref{eq:hazard}) this ratio can be expressed as a logistic probability:
\begin{eqnarray*}
	 \frac{\lambda_{sr}^{(0,1)}(t_k)}{\lambda_{sr}^{(0,1)}(t_k)+\lambda_{sr}^{(1,0)}(t_k)} &=& 
	 \frac{\exp{\Big( \Delta_k  + 
			(\beta^{(0,1)}- \beta^{(1,0)})^\top x_{sr}(t_k)\Big)}}{1+\exp{\Big( \Delta_k  + 
			(\beta^{(0,1)}- \beta^{(1,0)})^\top x_{sr}(t_k)\Big)}}\\
	 	&=&   \frac{\exp(\Delta_k  +\Delta\beta^\top  x_{sr}(t_k) )}{1+\exp(\Delta_k  +\Delta\beta^\top  x_{sr}(t_k) )}
\end{eqnarray*}
where $\Delta_k = \log(\lambda_{sr}^{(0,1)}(t_k)) - \log(\lambda_{sr}^{(1,0)}(t_k))$ and $\Delta\beta = \beta^{(0,1)}-\beta^{(1,0)}$ are the effect contrasts. The remaining exponential factor $1 - \exp\Big[-(\lambda_{sr}^{(0,1)}(t_k)+\lambda_{sr}^{(1,0)}(t_k))\Delta t_k\Big]$
represents the probability that some transition occurs within the time window $\Delta t_k$, we refer to it as an inertia term. Small values indicate strong persistence, high inertia, whereas larger values imply a low inertia and a higher rate of transitioning. We use the approximation that inertia depends mainly on length of each interval $\Delta t_k$ and the baseline rates $\lambda_0^{(0,1)}(t)$ and $\lambda_0^{(1,0)}(t)$ in that interval, whereas it is approximately constant across all pairs $(s,r)$ within an interval. Thus, for some $\rho_k < 0$, we have, approximately, that for all ties $(s,r)$,
\begin{equation*}
\exp(\rho_k) \approx 	1 - \exp\Big[-(\lambda_{sr}^{(0,1)}(t_k)+\lambda_{sr}^{(1,0)}(t_k))\Delta t_k\Big].
\end{equation*}
With the above adjustments, we can now write an approximation of each of the four terms in the full likelihood \eqref{eq:lik}.
\begin{enumerate}
	\item The first term in the full likelihood $L(\beta)$ can be written as
	\begin{align*}
		\frac{\exp(\Delta_k+\Delta\beta x_{sr}(t_k))}{1+\exp(\Delta_k+\Delta\beta^\top x_{sr}(t_k))} \cdot \exp(\rho_k)  =& \frac{\exp(\Delta_k+\Delta\beta^\top x_{sr}(k) + \rho_k)}{1+\exp(\Delta_k+\Delta\beta^\top x_{sr}(k))} \\=& \frac{\exp(\Delta_k+\Delta\beta^\top x_{sr}(k) + \rho^*_k)}{1+\exp(\Delta_k+\Delta\beta^\top x_{sr}(k)+\rho^*_k)},
	\end{align*}
	for some $\rho^*_k < \rho_k <0$ acting as an equivalent shift on the log-odds scale that absorbs the inertia effect. 
	\item Accordingly, the second term in the full likelihood becomes 
	\begin{align*}
		1-\frac{\exp(\Delta_k+\Delta\beta^\top x_{sr}(k) + \rho^*_k)}{1+\exp(\Delta_k+\Delta\beta^\top x_{sr}(k)+\rho^*_k)} = \frac{1}{1+\exp(\Delta_k+\Delta\beta^\top x_{sr}(k)+\rho^*_k)}.
	\end{align*}
	\item Note that the third term simply exchanges the roles of the states compared to the first term and can therefore be written as 
	\begin{align*}
		\frac{\exp(-\Delta_k-\Delta\beta^\top x_{sr}(k))}{1+\exp(-\Delta_k-\Delta\beta^\top x_{sr}(k))} \cdot \exp(\rho_k)  = 
		& \frac{\exp(-\Delta_k-\Delta\beta^\top x_{sr}(k) + \rho^*_k)}{1+\exp(-\Delta_k-\Delta\beta^\top x_{sr}(k)+\rho^*_k)}  \\
		=&\frac{1}{1+\exp(\Delta_k+\Delta\beta^\top x_{sr}(k)-\rho^*_k)},
	\end{align*}
	\item and consequently the fourth term can be written as
	\begin{align*}
		1 - \frac{1}{1+\exp(\Delta_k+\Delta\beta^\top x_{sr}(k)-\rho^*_k)} =  \frac{\exp(\Delta_k+\Delta\beta^\top x_{sr}(k)-\rho^*_k)}{1+\exp(\Delta_k+\Delta\beta^\top x_{sr}(k)-\rho^*_k)}.
	\end{align*}
\end{enumerate}  
Using these approximations, the full likelihood is approximated using the following logistic likelihood,
\begin{align*}
	L(\boldsymbol{\beta},\boldsymbol{\rho},\boldsymbol{\Delta})  \approx \prod_{k=1}^{n-1} 	\prod_{sr} &\Bigg[  \frac{\exp(\Delta_k+\Delta\beta^\top x_{sr}(k)+\rho^*_k z_{sr}(k))}{1+\exp(\Delta_k+\Delta\beta^\top x_{sr}(k)+\rho^*_kz_{sr}(k))}\Bigg]^{y_{sr}(k)} \times \\
	&\Bigg[  \frac{1}{1+\exp(\Delta_k+\Delta\beta^\top x_{sr}(k)+\rho^*_k z_{sr}(k))}\Bigg]^{1-y_{sr}(k)},
\end{align*}
where the dummy response $y$ and a dummy inertia covariate $z$ are defined as
\begin{align}
	y_{sr}(k) =& \begin{cases}
		1, & \text{ if } (s,r) \text{ is in state 1 at time } t_{k+1}, \\
		0, & \text{ otherwise},
	\end{cases}\\
	z_{sr}(k) =& \begin{cases}
		1, & \text{ if } (s,r) \text{ is in state 0 at time } t_{k}, \\
		-1, & \text{ otherwise}.
	\end{cases}
\end{align}
This approximation of the full likelihood enables estimation of the effects via standard logistic regression techniques. The baseline hazard contrast terms $\Delta_k$ can either be estimated discretely or via a smooth term in case sufficient number of panels are available. Similarly, the inertia parameters $\boldsymbol{\rho}$ can be estimated. Furthermore, only the effect contrasts $\Delta\beta$ are identifiable, unless one is willing to make SOAM-like restrictions, $\beta^{(0,1)}=-\beta^{(1,0)}$, as discussed in section~\ref{subsec:parameters}.

\section{Simulation study}

We evaluate the proposed framework in a series of simulation studies designed to assess its ability to recover known transition dynamics. In particular, we consider two data sampling scenarios: (i) full-history setting, in which every change of a dyad’s state is observed with its exact time; (ii) panel setting, in which only the network states at specific times are observed and all within-interval transitions are latent. The aim is to demonstrate that the framework recovers known patterns of network evolution under both information-rich and a more realistic, interval-censored designs.

\subsection{Full-history setting}

We begin with the case where the complete sequence of tie changes is observed. Consider a directed network of 20 nodes in which each edge occupy one of three states $\mathcal{I} = \{0,1,2\}$. Initial edge states are assigned at random, and transitions from a current state $i$ to a target state $j$ are governed by the following intensity function, defined with the anchor-and-pull constraints defined in section~\ref{subsec:parameters},
\begin{align*}
	\lambda_{sr}^{(i,j)}(t) = \exp\Big[
	& \beta_{\mathrm{out}}^{-} \mathrm{OutDeg}_{s}^{(i)}(t) + 
	\beta_{\mathrm{out}}^{+} \mathrm{OutDeg}_{s}^{(j)}(t) \\ 
	& + \beta_{\mathrm{rec}}^{-}\,\mathrm{Rec}_{sr}^{(i)}(t) + \beta_{\mathrm{rec}}^{+} \mathrm{Rec}_{sr}^{(j)}(t) \\ 
	&
	+ f^{-}\big(\mathrm{Rb}_{sr}^{(i)}(t)\big)
	+ f^{+}\big(\mathrm{Rb}_{sr}^{(j)}(t)\big)  \Big], \qquad i \neq j,
\end{align*}
where $\mathrm{Rec}_{sr}^{(i)}(t)$ denotes reciprocity in state $i$ at time $t$, $\mathrm{OutDeg}_{s}^{(i)}(t)$ is the sender node's out-degree in state $i$ at time $t$, and $\mathrm{Rb}_{sr}^{(i)}(t)$ is the receiving balance in state $i$ at time $t$, defined as the number of third nodes that maintain outgoing ties in state $i$ to both the sender $s$ and the receiver $r$:
\begin{equation*}
	\text{Rb}^{(i)}_{sr}(t)  = \sum_{h \notin \{s,r\}} \mathbf{1}_{\{ E_{hs}(t) =i= E_{hr}(t)\}}.
\end{equation*} 
The true effects are set to
$\beta_{\mathrm{out}}^{-} = -2, \beta_{\mathrm{out}}^{+} = 2,\beta_{\mathrm{rec}}^{-} = -1, \beta_{\mathrm{rec}}^{+} = 1$,
and the receiving balance effects enter through logarithmic functions:
\begin{align*} f^{-}\big(\mathrm{Rb}_{sr}^{(i)}(t)\big) &= -0.5 \log(\mathrm{Rb}_{sr}^{(i)}(t) + 1),\\
	f^{+}\big(\mathrm{Rb}_{sr}^{(j)}(t)\big) &= -\log(\mathrm{Rb}_{sr}^{(j)}(t) + 1),
\end{align*}
so that increases in receiving balance reduce intensities with diminishing returns, and the reduction is stronger for the pulling component than for the anchoring component.

\begin{figure}[t]
	\centering
	\subfloat[\label{fig:fullres-i}]{
		\includegraphics[width=.5\textwidth]{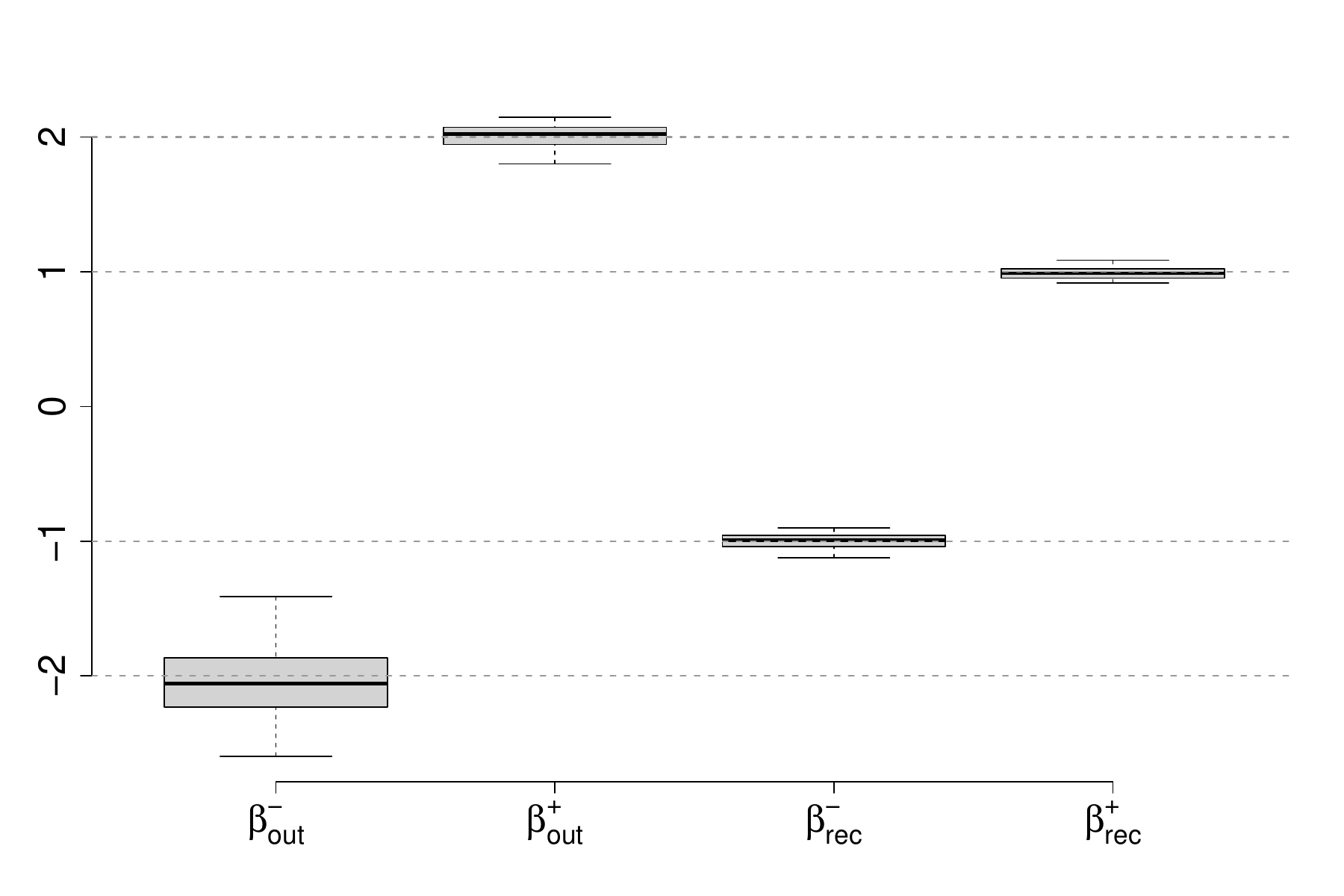}
	}
	\subfloat[\label{fig:fullres-j}]{
		\includegraphics[width=.5\textwidth]{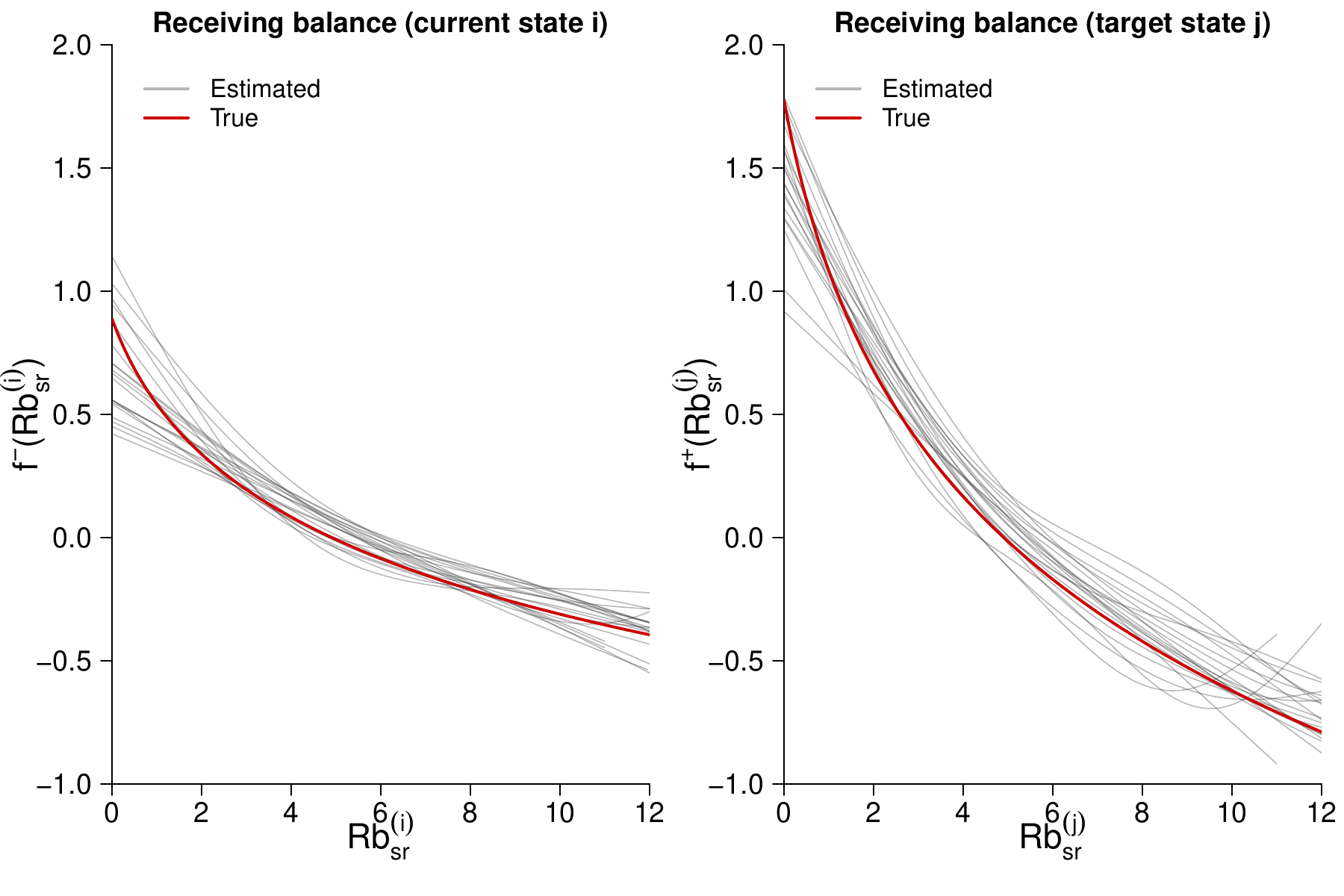}
	}
	\caption{Estimates in the full-history setting. (a) Boxplots of the estimated linear effects show that estimates are centred on the true values (grey reference lines) with low dispersion. (b) Estimated non-linear receiving balance effects for current state $i$ (left) and target state $j$ (right) closely follow the true decreasing, concave curves.}
	\label{fig:fullres}
\end{figure}

In accordance with this scenario, for each replication we simulate 10,000 state changes and estimate the model using the full transition history. Figure \ref{fig:fullres} summarises results over 20 replications. Panel \ref{fig:fullres-i} reports boxplots of the estimated linear effects. Across replications, estimates recover the true magnitudes with low dispersion, indicating minimal bias relative to sampling variability. Panel \ref{fig:fullres-j} displays the non-linear receiving balance effects for the current state $i$ and the target state $j$. The estimated curves closely follow the true functions, reproducing the expected decreasing, concave shapes. Moreover, the pulling effect $f^{+}$ is consistently steeper than the anchoring effect $f^{-}$, as designed $-\log(\cdot + 1)$ vs. $-0.5 \log(\cdot + 1)$. Overall, the figure demonstrates accurate recovery of both linear coefficients and non-linear state-dependent effects when the full event history is observed.

\subsection{Panel-data setting}

We now consider the more common panel-data setting, in which the network is observed only at discrete time points, producing cross-sectional snapshots. The underlying process still evolves continuously and may undergo multiple tie changes between observations. However, only the states at the observation times are recorded, so all within-interval transitions are latent. 

We focus on a two-sate model, $\mathcal{I} = \{0,1\}$ and simulate a directed network with 20 nodes. Initial edge states are assigned at random, and subsequent evolution follows state-dependent transition intensities. In the two-states case, many state-specific covariates are linearly dependent. For instance, for any sender $s$, 
\begin{equation*}
	\mathrm{OutDeg}_s^{(1)}(t)= (p-1)-\mathrm{OutDeg}_s^{(0)}(t),
\end{equation*}
where $p$ is the number of nodes. Accordingly, we define the transition intensities using only state 0 covariates and allow coefficients to differ by direction of change,
\begin{align*}
	&\lambda_{sr}^{(0,1)}(t) = \exp\Big[
	\beta_{\mathrm{out}}^{-} \mathrm{OutDeg}_{s}^{(0)}(t)
	+\beta_{\mathrm{rec}}^{-}\,\mathrm{Rec}_{sr}^{(0)}(t) 
	+ \beta_{\mathrm{rb}}^{-}  \log(\mathrm{Rb}_{sr}^{(0)}(t) + 1) \Big]\\
	&\lambda_{sr}^{(1,0)}(t) = \exp\Big[
	\beta_{\mathrm{out}}^{+} \mathrm{OutDeg}_{s}^{(0)}(t)
	+ \beta_{\mathrm{rec}}^{+}\,\mathrm{Rec}_{sr}^{(0)}(t) 
	+ \beta_{\mathrm{rb}}^{+}  \log(\mathrm{Rb}_{sr}^{(0)}(t) + 1)\Big],
\end{align*}
where $\mathrm{Rec}_{sr}^{(0)}(t)$ indicates wheter the reverse tie $(r,s)$ is in state 0 at time $t$, $\mathrm{OutDeg}_s^{(0)}(t)$ is the out-degree of senders $s$ in terms of state 0, and $\mathrm{Rb}_{sr}^{(0)}(t)$ is the receiving balance in state 0. The superscripts “$-$” and “$+$” on the coefficients distinguish anchoring effects (governing moves away from state 0 in the $0\to1$ transition) from pulling effects (governing moves toward state 0 in the $1\to0$ transition). Under this parametrisation, the same covariate may play opposite roles across directions. For example, $\mathrm{Rec}_{sr}^{(0)}(t)$ may discourage movement $0\to1$ via $\beta_{\mathrm{rec}}^{-}$ while encouraging movement from $1\to0$ via $\beta_{\mathrm{rec}}^{+}$. 

The data-generating parameters are set to $\beta_{\mathrm{out}}^{-} = 1, \beta_{\mathrm{out}}^{+} = 2$ for out-degree, $\beta_{\mathrm{rec}}^{-} = 2, \beta_{\mathrm{rec}}^{+} = 1$ for reciprocity, and $\beta_{\mathrm{rb}}^{-} = -2,   \beta_{\mathrm{rb}}^{+} = -3$ for receiver balance.
Given the intensity rates, we simulate 10,000 state changes. We partition the observation horizon into five equal length intervals and record a network snapshot at each interval boundary. Using these panel snapshots, we compute the required network statistics and estimate the model using approach described in section~\ref{subsubsec:two_state}.

\begin{figure}
	\centering
	\includegraphics[width=0.7\linewidth]{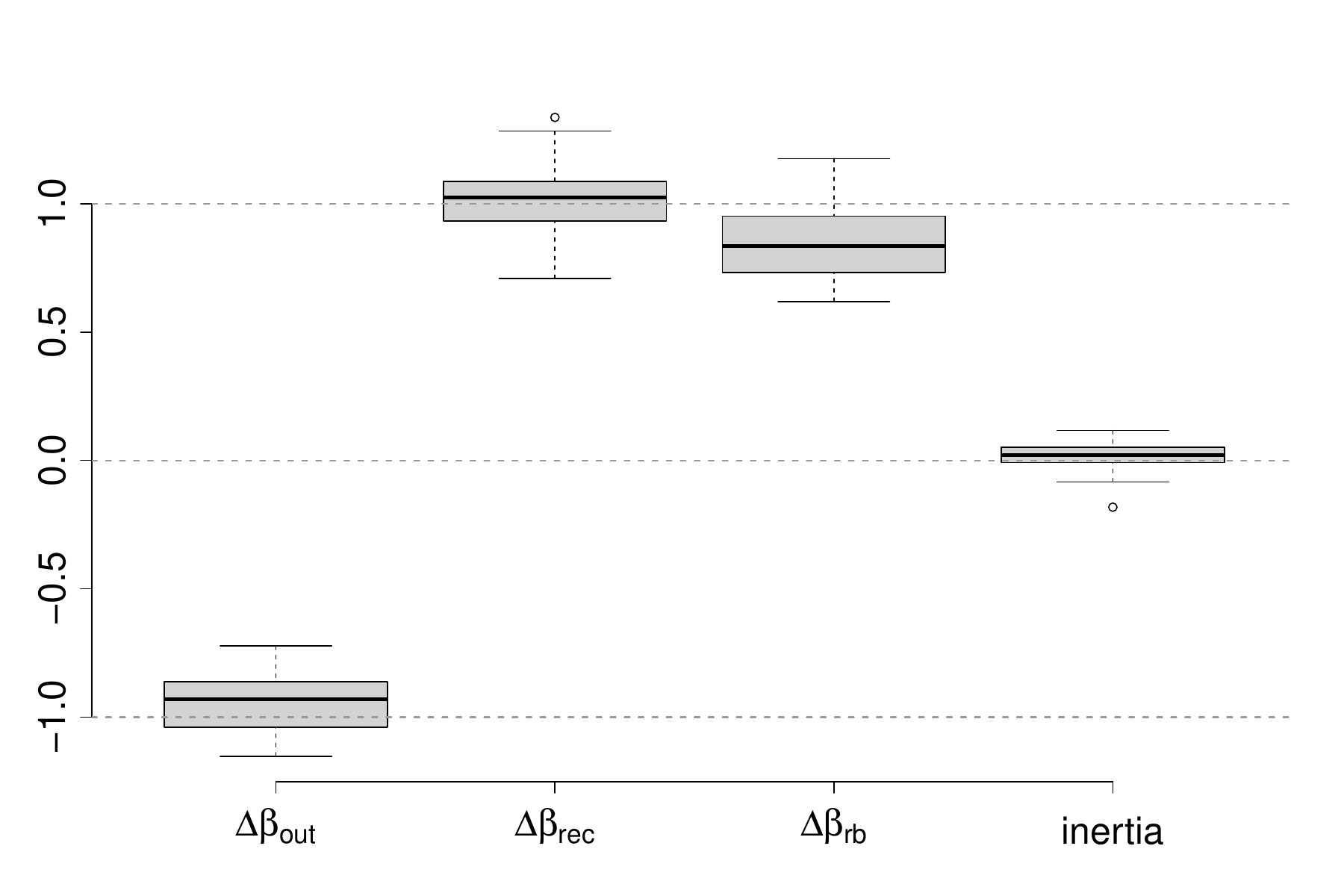}
	\caption{Estimates in the panel data setting. Boxplots of the estimated contrasts are centred near the true values (grey reference lines) with modest dispersion, recovering the negative contrast for out-degree and the positive contrasts for reciprocity and receiving balance.}
	\label{fig:panelres}
\end{figure}

Figure~\ref{fig:panelres} summarises results over 20 replications. Because formation and dissolution coefficients are not separately identifiable from panel data without additional restrictions, we report estimates of the contrasts $\Delta\beta = \beta^{-} - \beta^{+}$ along with the inertia parameter. The contrasts are well centred on their true values with modest dispersion; small biases in some terms are attributable to the inertia approximation. The inertia parameter is estimated as constant across intervals and is close to zero, as expected given the high volume of changes relative to the temporal resolution of the five-interval panel. Overall, these simulations show that the proposed estimator recovers effect contrasts reliably, even when absolute coefficients cannot be disentangled.

\section{Teenage Friends and Lifestyle Study}


We apply the novel inference procedure to the Teenage Friends and Lifestyle Study \cite{bush1997role,michell1996peer}, which followed a cohort of pupils in West of Scotland over a three–year period from 1995 to 1997. At each wave, pupils could nominate up to twelve friends, producing three panel snapshots of the evolving friendship network. A total of 160 pupils participated in the study. Alongside network data, information was collected on a set of behavioural variables, including tobacco, alcohol, and cannabis usage. Behavioural variables were measured on ordered categorical scales: smoking (1 = non-smoker, 2 = occasional smoker, 3 = regular smoker, i.e. more than once per week), cannabis use (1 = non-user, 2 = tried once, 3 = occasional user, 4 = regular user), and alcohol consumption (1 = non-drinker, 2 = once or twice a year, 3 = once a month, 4 = once a week, 5 = more than once a week). In addition to these exogenous covariates, we incorporated a set of endogenous network effects to capture structural dependencies that are commonly observed in social networks. Specifically, we considered reciprocity (the tendency for friendship nominations to be reciprocated), out-degree (the general inclination to maintain a larger number of outgoing ties), and distance-2 effect (the influence of indirect connections, measured by the number of actors reachable within two steps). We further accounted for homophily and sender/receiver effects for gender and substance use attributes. 

Besides considering an empirical application, this analysis will also allow for an empirical validation of the proposed inference procedure by comparing it directly to output of a SAOM estimation procedure.  For panel data, the SAOM has become the standard tool for analysing longitudinal network data. SAOM models are limited to two-state networks, where ties are either present or absent. To enable a fair comparison, we therefore restrict attention to the two–state scenario introduced in section~\ref{subsubsec:two_state} and evaluate both approaches side-by-side. 

To ensure comparability, we specified covariates in our framework following the same definitions as in the published SAOM analysis, where they determine the change in utility associated with a potential tie change. Moreover, SAOM imposes restrictions on coefficients: in particular, tie formation and dissolution are constrained to have anti-symmetric effects, as define in section~\ref{subsec:parameters}. By adopting identical specifications and imposing the same constraints, we ensure that both models capture the same mechanisms and are directly comparable.

Parameter estimates from our framework are compared with those obtained from the SAOM \cite{pearson2006homophily} in Table~\ref{tab:glasgow_results}. Qualitatively, the two approaches yield very similar findings. Both identify strong and significant effects of reciprocity, out-degree, and distance-2, as well as a consistent tendency toward gender homophily. Both models suggest that pupils tend to reciprocate friendships, the negative out-degree effect indicates they are not inclined to maintain large numbers of ties, the negative distance-2 effect suggests a preference for direct rather than indirect connections, and gender homophily confirms a tendency to nominate same-gender friends. Differences between the models appear mainly in the magnitude of the coefficients and in the significance of some behavioural covariates, but the overall substantive conclusions remain the same. Our framework also yields a wave-specific inertia parameter, which captures the baseline tendency of ties to persist over an observation interval. This parameter is inversely related to the SAOM rate parameter: higher SAOM rates correspond to lower inertia. Both of these measures indicate that friendship tie formation is somewhat lower in the second observation wave. 

Despite the quantitative similarity of the results, the proposed inference procedure achieves a far greater computational efficiency, with estimation taking only 0.26 seconds. By contrast, although exact timings for this particular SAOM specification are unavailable, related work has shown that estimating an extended SAOM for the same dataset (with 52 parameters) required 39 hours using RSiena \cite{steglich2006applying}. Since our framework relies on logistic regression, estimation time increases only modestly with additional covariates, making our approach far more scalable to larger or more complex specifications.

\begin{table}[bt]
	\centering
	\begin{tabular}{lrrrr}
		\toprule
		& \multicolumn{2}{c}{New approach} & \multicolumn{2}{c}{SAOM}\\
		Parameter & Estimate & p-value& Estimate & p-value \\
		\midrule
		distance-2 & -0.92 & $<0.001 $& -1.08 &  $<0.001 $\\
		out-degree   & -1.81 & $<0.001$& -1.98 &  $<0.001 $\\
		reciprocity   & 2.22  & $<0.001$& 2.29&  $<0.001 $\\
		gender homophily    & 0.66  & $<0.001$& 0.78&  $<0.001 $ \\
		sender female           & -0.06 & 0.48 & 0.12 & 0.32 \\
		receiver female          & 0.07  & 0.4 & -0.17 & 0.19\\
		tobacco homophily  & -0.16 & 0.11&  0.42 & 0.22\\
		sender tobacco      & 0.05  & 0.62 & 0.28&  0.03 \\
		receiver tobacco    & -0.23 & 0.03 & -0.25& 0.05\\
		cannabis homophily   & -0.2 & 0.15 & 0.18 & 0.72 \\
		sender cannabis     & 0.72  &  $<0.001$ &  -0.15 & 0.10 \\
		receiver cannabis   & -0.06 & 0.82 &  0.09 & 0.37\\
		alcohol homophily   & 0.21  & 0.01 & 0.96 & 0.01\\
		sender alcohol      & 0.05 & 0.67 & -0.04& 0.32\\
		receiver alcohol    & 0.29  & 0.01 & 0.06& 0.23\\
		inertia period 1       & -0.06 & 0.53 & - & -\\
		inertia period 2      & -0.20 &  0.04 & - & -\\
		rate period 1       & - & -& 12.72 & -\\
		rate period 2      & - & -& 9.65 & -\\
		\bottomrule
	\end{tabular}
	\caption{Comparison of parameter estimates from the proposed framework and SAOM on the Teenage Friends and Lifestyle Study.}
	\label{tab:glasgow_results}
\end{table}

\section{Conclusions}

This paper introduced a continuous‐time framework for modelling multivariate relational states in dynamic networks, extending binary tie models to settings where each edge occupies one of several substantively distinct states and evolves by transitioning between them over time. Transition intensities are allowed to depend on state‐dependent covariates that might act as anchoring (current–state) and pulling (target–state) mechanisms, with both linear and non-linear components accommodated within a generalized additive formulation. This yields a flexible specification that captures how various endogenous network mechanisms (e.g., reciprocity, degree, triadic closure) along with exogenous covariates operate across different stages of tie evolution.

Methodologically, we provided estimation strategies for two common observation schemes. For full event histories, a partial‐likelihood approach with efficient nested case–control sampling recovers both linear coefficients and smooth effects with high precision. For panel snapshots, we derived an approximation to the full likelihood over short intervals; however, the aggregation of many competing processes complicates estimation, and the analysis of general multi-state panel case still remains an open question. In the two-state care, it is possible to construct a likelihood approximation that factorises into a logistic component for the relative propensity of tie formation versus dissolution, and an inertia component capturing baseline persistence between waves. In this design, absolute formation and dissolution coefficients are not separately identifiable; instead, their contrasts are estimable, which is sufficient for comparing the direction and relative strength of covariate effects. When additional restrictions are imposed (e.g., SAOM style symmetry restriction), the  coefficients are identifiable and can be estimated.

Simulation studies support the efficiency of the proposed approach. In the full‐history setting, the estimator recovers the true magnitudes and non-linear shapes with low dispersion across replications. In the panel setting, the identifiable contrasts are well centred on the truth, confirming that meaningful inference remains possible when within‐interval events are latent. An empirical application to the Teenage Friends and Lifestyle Study shows that the proposed approach reproduces the substantive conclusions of SAOM under matched specifications while offering substantial computational gains. Because estimation reduces to logistic regression fitting, runtime scales in a cubic manner with the number of covariates, quadratically with the number of individuals in the study, and linearly with the number of panels, making the method attractive for larger or more complex designs.

Several limitations suggest directions for future work. First, the panel estimator identifies contrasts rather than absolute coefficients; more sophisticated identification strategies or post-hoc analysis might recover the original effects. Secondly, we used a very parsimonious approximation for inertia. Richer persistence structures, either state-, node-, or dyad-specific, merit further investigation. Finally, extending the panel likelihood beyond the two‐state case with scalable approximations to matrix exponentials would bring the full multi‐state model within reach for snapshot data. In sum, the proposed framework provides a scalable approach to analysing persistent relational states. It retains the interpretability of classical network effects, generalises them to multi‐state ties, and enables estimation under both ideal (full‐history) and more common (panel) observation schemes, thereby broadening the empirical scope of dynamic network analysis.


\bibliographystyle{plain}
\bibliography{bibliography}
\end{document}